# Bus transport network analysis in Rio de Janeiro based on topological models using Social Networks

Louise Pumar[1], Rafael Barbastefano[1] and Diego Carvalho[1]

[1] Centro Federal de Educação Tecnológica Celso Suckow da Fonseca, Rio de Janeiro, Brazil
corresponding author: d.carvalho@ieee.org

**Abstract.** In recent years, studies on public transport networks have intensified in the Social Networks' field, especially in bus networks. This has occurred because of the relevance of urban mobility for the proper functioning of a city. The Rio de Janeiro city, Brazil, has undergone recent changes in the city of Rio de Janeiro, Brazil, has undergone recent changes in its municipal bus system, modifying several lines and bus stops. This paper proposes to analyze the structure of the bus transportation network of this city, comparing its topology in 2014 and 2016 – before and after the change. For this, the properties of the bus system were investigated based on the topological models B-space, P-space and C-space. Some important parameters were calculated, such as giant component, distance, diameter, degree, closeness and betweenness. The results showed that there was a reduction of 22.75% of the lines and 5.19% of the bus stops from 2014 to 2016. It was also verified that a maximum of four lines are required to move between any two bus stops within the city in both years. However, with three lines is possible to reach more than 99% of the bus stops. Besides, this study also suggests exploring the C-space network according to a minimum number of common bus stops that the lines had. Based on the component giant analysis of these networks with many common points, it is possible to detect possible expressway corridors.

## 1 Introduction

Social network analysis has attracted considerable interest and curiosity from the social and behavioral science community in recent decades. Much of this interest can be attributed to the appealing focus of social network analysis on relationships among social entities, and on the patterns and implications of these relationships [1].

One of the enforcement fields of social networks is urban mobility, since this tool allows the construction and analysis of transport networks. Urban mobility is paramount for the proper functioning of a city since the population needs to travel daily to carry out its activities, especially regarding home-work relationship. With a growing population, the transportation demand also increases, causing more traffic on the streets. Consequently, this growing creates more problems related to mobility, such as traffic jams, air pollution, noise pollution and accidents. It occurs mostly in urban centers where the level of human activities is high [2].

Due to the importance of mobility and the various problems caused by it, many cities – especially larger cities – have begun investing in public transport development, encouraging and improving their structure in order to make them more attractive, faster and efficient. As the public transport of the city can be arranged in a network with significant amounts of nodes and links, several studies with social networks applications in public transport networks have begun to emerge.

This study aims to apply topological models used in social networks to the public transport network, specifically evaluating the structure of the bus system in the city of Rio de Janeiro. The routes and bus stops of this city have undergone recent modifications. So, this article aims to compare the network topology before and after this change, comparing the properties analyzed in the networks in 2014 and 2016.

In addition, this study also proposes a modification in the one of the topological models application used, creating different C-space networks according to the number of bus stops that routes had in common.

After this introduction, the present article is structured as follows: Section 2 presents the topological models used, highlighting some works where these models were applied. In Section 3 we find the properties used in network analysis. A contextualization about public transport in the city of Rio de Janeiro is presented in Section 4. The proposed methodology is shown in Section 5, while the results of this modeling are found in Section 6. Final considerations and recommendations for the pursuance of this research are presented in the last section.

## 2 PTN topological models

Von Ferber et al. [3] developed four topological models – L-space, B-space, P-space and C-space – to analyze the public transport networks (PTN) of certain cities. For this purpose, the authors treated the nodes of the networks as routes and/or bus stops, depending on the need of the analysis.

In the L-space topological model, the nodes of the network represent bus stops. An edge between the nodes exists if these nodes are adjacent in a given sequence [3]. In that case, the sequence is the chronological order of the points which a bus of a certain route will stop. Although this model is quite applied in PTN analysis, the data obtained in this study for the creation of networks did not allow the application of this model. It happened because the data has not defined a single and correct sequence of the bus stops in each route.

The B-space topological model is a two-mode network. According to Wasserman e Faust [1], a bipartite graph is defined as a graph where the nodes can be partitioned into two subsets and all lines are between nodes belonging to different subsets. Nodes in a given subset are adjacent to nodes from the other subset, but no node is adjacent to any node in its own subset.

In B-space, both routes and stations are represented by nodes, each being considered as a partition. Each route node is linked to all station nodes that it serves. As previously mentioned, no direct links between nodes of the same type (route or bus stop) occur. Obviously, in the B-space the neighbors of a given route node are all stations that it service while the neighbors of a given station node are all routes that service it [3].

P-space is one of the networks obtained from B-space, since it is the two-mode network transformed just to a partition – in this case, the bus stops partition. In this topological model, the nodes are stations and they are linked if they are serviced by at least one common route [3].

The topological model C-space is the other network obtained from the B-space, based on the routes partition. It contains route nodes and any two routes nodes are neighbors if they share a common station [3].

Fig. 1 shows an example of networks generated from these topological models. In this example, three different bus routes were created. Route A passes through points 1, 2, 3, 4 and 5. Route B contains points 2, 3, 5, 6, 7 and 8. Finally, route C uses points 3, 4, 5, 8, 9 and 10. Following, topological model networks based on B-space, C-space and P-space were generated.

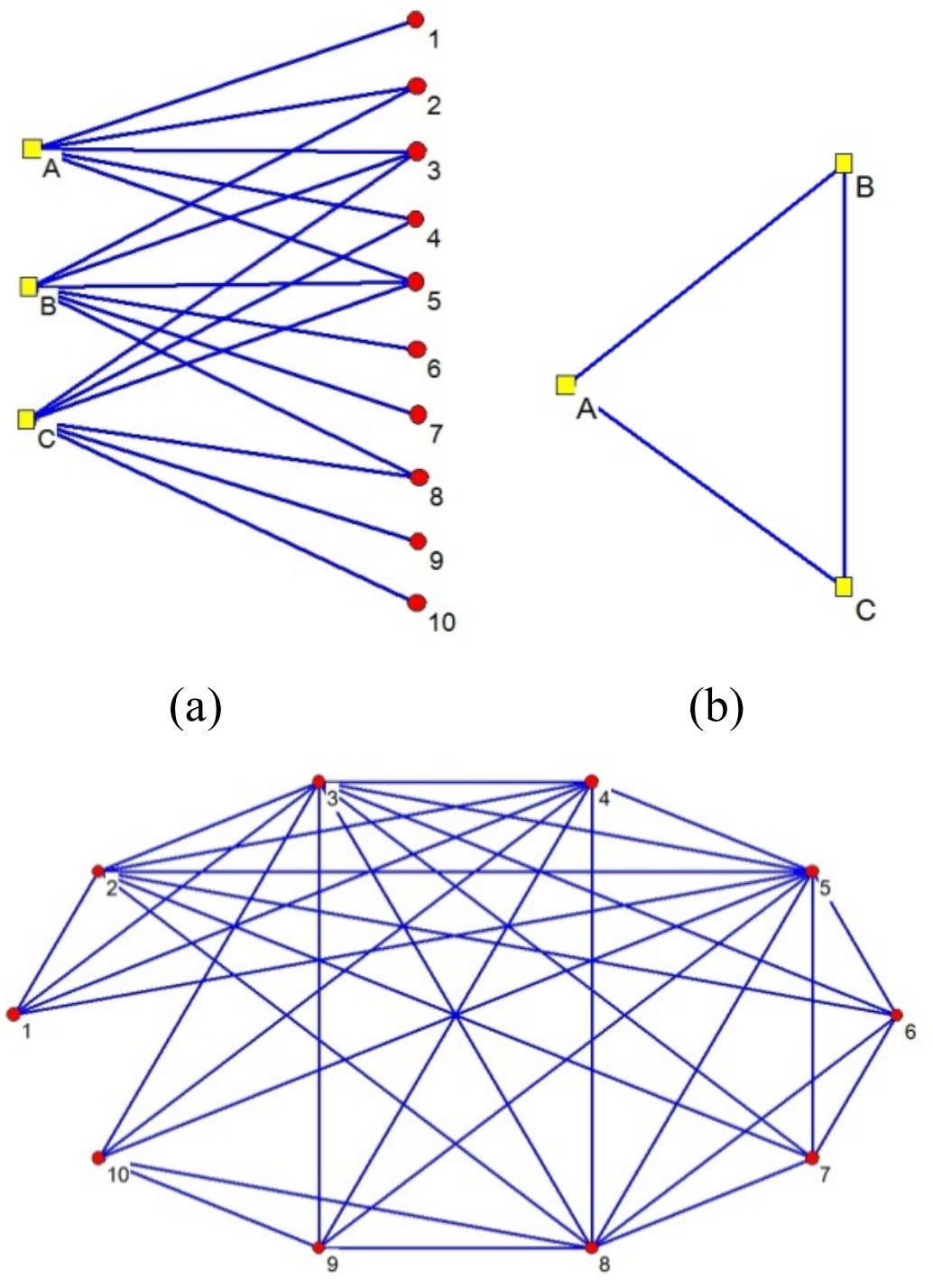

(c)

**Fig. 1.** (a) B-space. (b) C-space. (c) P-space

In order to facilitate, when we mention L-space, B-space, C-space and P-space networks in this work, we will point them out as Ls, Bs, Cs and Ps, respectively.

Some researchers have studied features of the PTN topology using social networks. Sen et al. [4] investigated the structural properties of the Indian rail network in order to verify if this network also behaved like most social networks. For such, stations were considered as nodes and an arbitrary pair of stations was said to be connected by a link when at least one train stopped at both stations. Thereby, the authors verified that this network behaved like a complex network. This study was considered a differential for the application of the concepts of social networks to PTN.

In 2009, von Ferber et al. [3] established four topological models of great utility for the PTN analysis: Ls, Bs, Ps and Cs, previously explained. According to the authors, the work had two main objectives. First, they wanted to present a systematic survey of statistical PTNs based on the data for cities of so far unexplored network size. The second objective was to present a model that with a small number of simple rules would be capable to reproduce the main properties. After this article publication, the topological models developed began to be studied by several authors and applied in different modal split of public transportation.

The topological models were mostly applied to the modal split of bus. Yu et al. [5], Wang and Yang [6], Zhang et al. [7] and Yang et al. [8] carried out case studies applying the topological models Ls and Ps to BTN for specific cities. Yang et al. [9], Xu et al. [10], Zhang et al. [11] and Feng et al. [12] also analyzed bus networks, but only used Ps for this, whereas Guo et al. [13] employed only Ls in their analysis. Zhang et al. [14] and An, Zhang and Zhang [15] performed more thorough investigations in their BTN studies, using three topological models in their works, Ls, Ps and Cs.

The topological modals were also applied to others modal split. Barberillo and Saldaña [16] studied the subway network of four major cities – New York, Paris, Barcelona and Moscow – while Liu and Tan [17] and Zhang, Zhang and Qiao [18] studied the subway network of Chinese cities. These three studies were performed using the topological models Ls and Ps. Qiao, Zhao and Yao [19] studied the train network of Beijing, China.

We also checked PTN studies with more than one modal split. Zou et al. [20] used the Ls and Ps topological models to analyze the railway and bus networks together. Yu, Ma e Zhang [21] applied these same topological models for researching the bus and subway networks in Nanjing, China.

## 3 Network properties

The social networks have several properties in order to facilitate their analysis and understanding. In this work, we studied and applied some of them to analyze the studied networks.

### 3.1 Degree

The degree of a node is defined as the number of edges that are incident with it. In other word, it is the number of nodes adjacent to it [1]. As higher is the node degree, more nodes are linked to it. This means that node degree reflects its importance as a hub [8]. Therefore, if a certain node has a high degree, it will probably have more visibility inside the network. It should be recognized as a main channel of relational information, occupying a central location. In contrast, low-level actors have a more peripheral position in the network [1].

Considering a graph *G* with *m* edges and *n* nodes, the average degree *dm(G)* is defined by:

$$dm(G) = (2 \times m) / n \tag{1}$$

### 3.2 Giant component

In networks with very small numbers of connections between individuals, all individuals belong only to small islands of collaboration or communication. As the total number of connections increases, however, there comes a point at which a giant component forms – a large group of individuals who are all connected to one another by paths of intermediate acquaintances [22]. Therefore, we can say that giant component is the component that contains most nodes in the network, since all of them are connected.

### 3.3 Distance

The geodesic distance between a pair of nodes in a graph is the length of a shortest path between the two nodes. In other words, it is the amount of minimum edges existing between these two nodes. If there is no path between two nodes (that is, they are not reachable), then the distance between them is infinite (or undefined) [1]. The distance between two nodes is the length of any path between them. It may not necessarily be the shortest path between these two nodes, but any path that connects them. In this work, in order to facilitate, when the distance property is mentioned, it will refer to the geodesic distance.

### 3.4 Diameter

Based on the distance definition described above, the diameter of a graph is the length of the largest geodesic distance between any pair of nodes in this graph [1]. Or more accurately, the diameter is the largest of these values found of all lengths values of the shortest paths between two nodes in a graph. The diameter is important because it quantifies how far apart the farthest two nodes in the graph are [1].

### 3.5 Closeness

Closeness is a centrality measure that focuses on how close an actor is to all the other actors in the set of actors. So, a node is central if it can quickly interact with all others. The main idea is that centrality is inversely related to distance. As a node grows farther apart in distance from other nodes, its centrality will decrease, since there will be more lines in the geodesics linking that node to the other nodes [1].

### 3.6 Betweenness

Betweenness is also a centrality measure. The important idea here is that an actor is central if it lies between other actors on their geodesics. This implies that to have a large betweenness centrality, the node must be between many of the nodes via their geodesics [1]. In other words, betweenness is based upon the frequency with a node falls between pairs of other nodes on the geodesic distance connecting them, considering all geodesic distance existing between them. This idea refers to maximum betweenness. In social networks, this measure is useful as an index of the potential of a actor for control of communication [23].

## 4 PT in the city of Rio de Janeiro

Public transportation in the city of Rio de Janeiro was initially developed by trains and tramways. Trains and tramways emerged in 1858 and 1868, respectively. They have brought to public transport a considerable increase in the number of passengers, giving them the mass transport proportion [24].

The first bus model was called auto-bus. It emerged in Rio de Janeiro in 1908, with the route Praça Mauá – Passeio Público. Soon other routes appeared, according to the population needs to move. In 1928, 19 companies were licensed, carrying over 88.1 million passengers per year [25].

In just over a decade, passenger traffic was signed. Bus stopped being a mass transit complement for the trains and tramways, and it developed into the main public transport mode. The bus became important for several reasons, since it allowed access to many places, without requiring rails. Also, it was faster and had a more flexible schedule than tramways and trains [25].

According to Juruá [26], during the 1960s and 1970s, Rio de Janeiro experienced a setback in mass urban transport – ferries, trains and subways. This was partly because the government made investments encouraging road transportation, such as Flamengo and Copacabana embankments, the construction of tunnels and viaducts and the construction of the Rio-Niterói bridge [25].

Currently, the city of Rio de Janeiro is the second largest city in Brazil, with about 6.5 million inhabitants in an area of 1,200 km$^2$ [27]. In Brazil, Rio de Janeiro is the main entry point for foreign tourists and also for investment, since it is the destination of great artistic and cultural performances and large-scale international sporting events. In addition, this city has the second largest Gross Domestic Product (GDP) in Brazil and it is the head office to many of the largest companies in the country [28].

PMUS [29] performed a study on the modal split used for the population displacements within the city of Rio de Janeiro. The results are presented in Fig. 2.

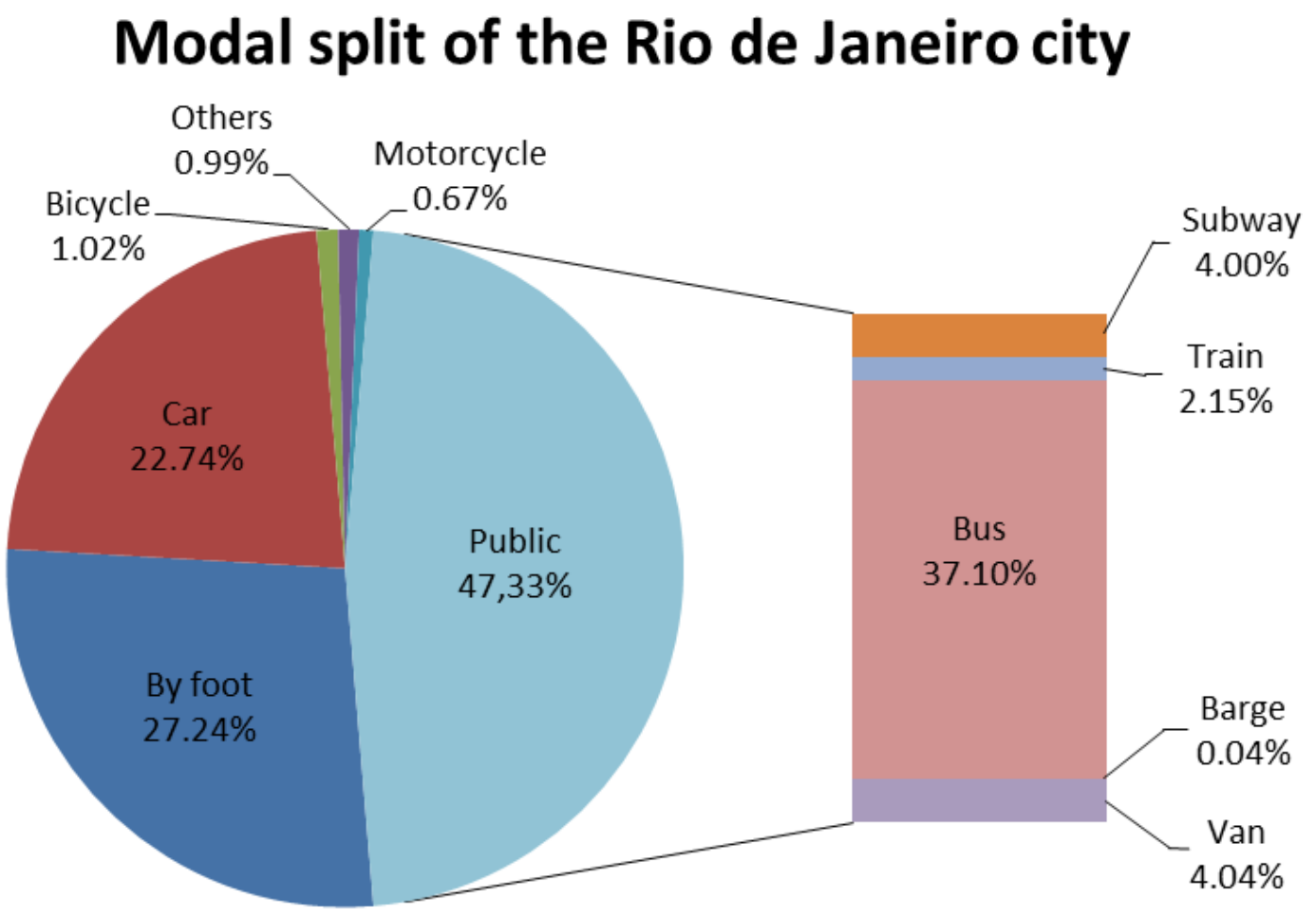

**Fig. 2.** Modal split of the Rio de Janeiro city. Source: Adapted from PMUS (2015)

Fig. 2 shows the relevance of public transport in this city, especially bus, since this was the modal split most used by the population, with 37.10% of the trips. According to Federação das Empresas de Transportes de Passageiros do Estado do Rio de Janeiro [30], the total number of conventional buses operating is 9,008 vehicles, with an average age of 4.38 years.

The Rio de Janeiro Government held a reorganization project of the municipal bus system in this city recently. This project began in 2010. However, only in 2015, the major changes in routes and bus stops were fulfilled [31].

This project contemplated two major motions: the rationalization of the South Zone bus system and the implementation of the Bus Rapid Transit (BRT), besides some small actions taken. The rationalization of the South Zone buses consisted in the introduction of Bus Rapid Service (BRS) in the main roads of this area, to replace many routes that passed through these roads, by adding bus lanes [31]. BRT is a large capacity articulated bus system, with buses larger than the conventional ones. It circulate in bus lanes, as well as it stops at all pre-established stations [32].

Due to this reorganization project, several routes and bus stops had to be modified or extinguished, bringing forth a major change in the city's bus system. It could be verified by comparing the bus network topology of 2014 with 2016. This work focused in this comparison.

## 5 Methodology

This work is a study case. Some BTN proprieties in the city of Rio de Janeiro were analyzed by three topological models – Bs, Ps and Cs. The main objective was the comparison and analysis of the BTN structure of this city in 2014 and 2016.

The first step was to collect the bus data of this city in its website [33]. The information was collected in two stages, one in 03/05/2014 and another in 29/01/2016. The website contains information about the municipal road transport provided by buses' GPS (Global Positioning System). This information is available in GTFS files and it covers all bus stops in the city, showing name and line's number, bus stops of each route, responsible agency for the route and latitude and longitude of bus stops.

Then, the data was processed. Since the necessary information was scattered in several files, a database was used to unify the information into a unique table. From this table, the data were formatted according to the required configuration by Bs topological model. It was transformed into a new file with NET format.

Next, the Bs network was opened in a specific program for network analysis. In this program, the Ps and Cs networks were obtained from the bipartite network Bs. Due to the networks complexity, we used the program called Pajek. This program was used because, according to Nooy et al. [34], it was developed to handle very large networks (up to millions of vertices), as it is the case of one created network – Ps.

At this stage, the Csn network was developed. This is a network created according to the number of bus stops that the routes have in common. To create this, we used the Cs network. The edges that connected two routes were removed when they had values lower than "n" common bus stops.

For a better understanding of Csn network, an example was developed. We considered five different bus routes. Line A contains bus stops 1, 2, 3, 4 and 5. Line B goes through the bus stops 2, 4, 6 and 8. Line C has points 2, 6 and 7. Line D, points 2, 3 and 4. And line E uses bus stops 1, 4, 6 and 7. From these routes, the Cs was drawn up, by connecting lines that have any common bus stops.

The Fig. 3a presents the $Cs^1$ network, which is the original Cs. In this network, the routes are connected if they have at least one common bus stop. The Fig. 3b shows the $Cs^2$ network. At this one, the edges that connected lines with just one common bus stop were removed. The $Cs^3$ network (Fig. 3c) only routes with three or more common bus stops remained connected. In this example, only lines A and D had three common bus stops.

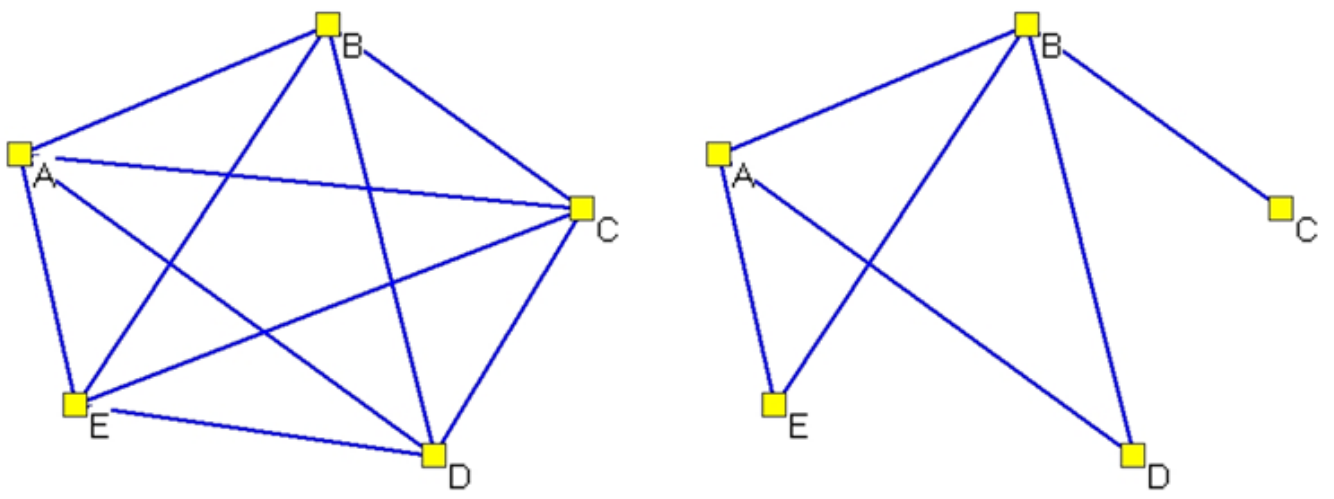

(a) (b)

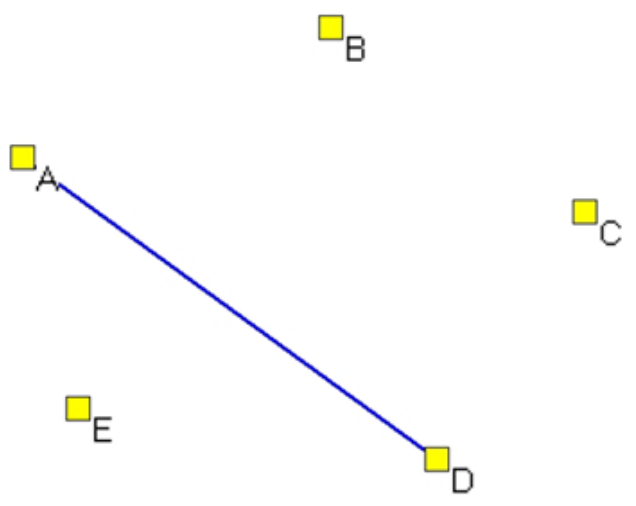

(c)

**Fig. 3.** Example of Csn network. (a) Cs1 network. (b) Cs2 network. (c) Cs3 network

In this work, Cs networks were created with 1 (which was the original Cs network), 50 and 100 bus stops in common. These were called Cs[1], Cs50 and Cs100, respectively.

Afterwards, in the Pajek program, we extracted several properties of the created networks. Because of their different topological model, the evaluated proprieties significance also diverged according to the network model. In general, the proprieties of degree, giant component, distance, diameter, closeness and betweenness were analyzed. These data were processed to compare and analyze proprieties values. This analysis allowed us to infer about the studied networks for a better understanding of the bus transport system structure of the Rio de Janeiro city, comparing the bus network before and after the change.

# 6 Results

According to the Cs[1] and Ps network nodes number, the Rio de Janeiro's BTN had 488 bus routes and 7020 bus stops in 2014. In 2016, it had 377 bus routes and 6656 bus stops. Therefore, there was a reduction of 22.75% in the routes amount and 5.19% in the bus stops amount. This shows that there was a large reduction in the routes number, most of them goes through stops where other lines were already passing. It can be inferred because the reduction of bus stops number was much lower.

The average distance in the Ps network shows how many bus routes are needed to get from one bus stop to another within the city, on average. In 2014, an average of 2.17 lines were necessary; in 2016, 2.22. That means that there was an increase in the routes number needed, on average, to go from one point to another. In the Ps network, the diameter represents the maximum routes amount necessary to move between the two most distant bus stops within the city. The diameter remained the same as 4 in 2014 and 2016.

Fig. 4 shows the distance distribution in the Ps network in 2014 and 2016, by comparing them. This graph presents the percentage of bus stops number that can be reached using a certain number of lines. Despite the bus system change, we can notice that the graphics are very similar. Although the networks' diameter is 4, it is possible to reach almost all existing bus stops in the city in both years (more than 99%) with a distance of 3. This means that passengers will need a maximum of 4 lines to move between the two most distant bus stops in the city, and with just 3 lines it is possible to reach almost all points.

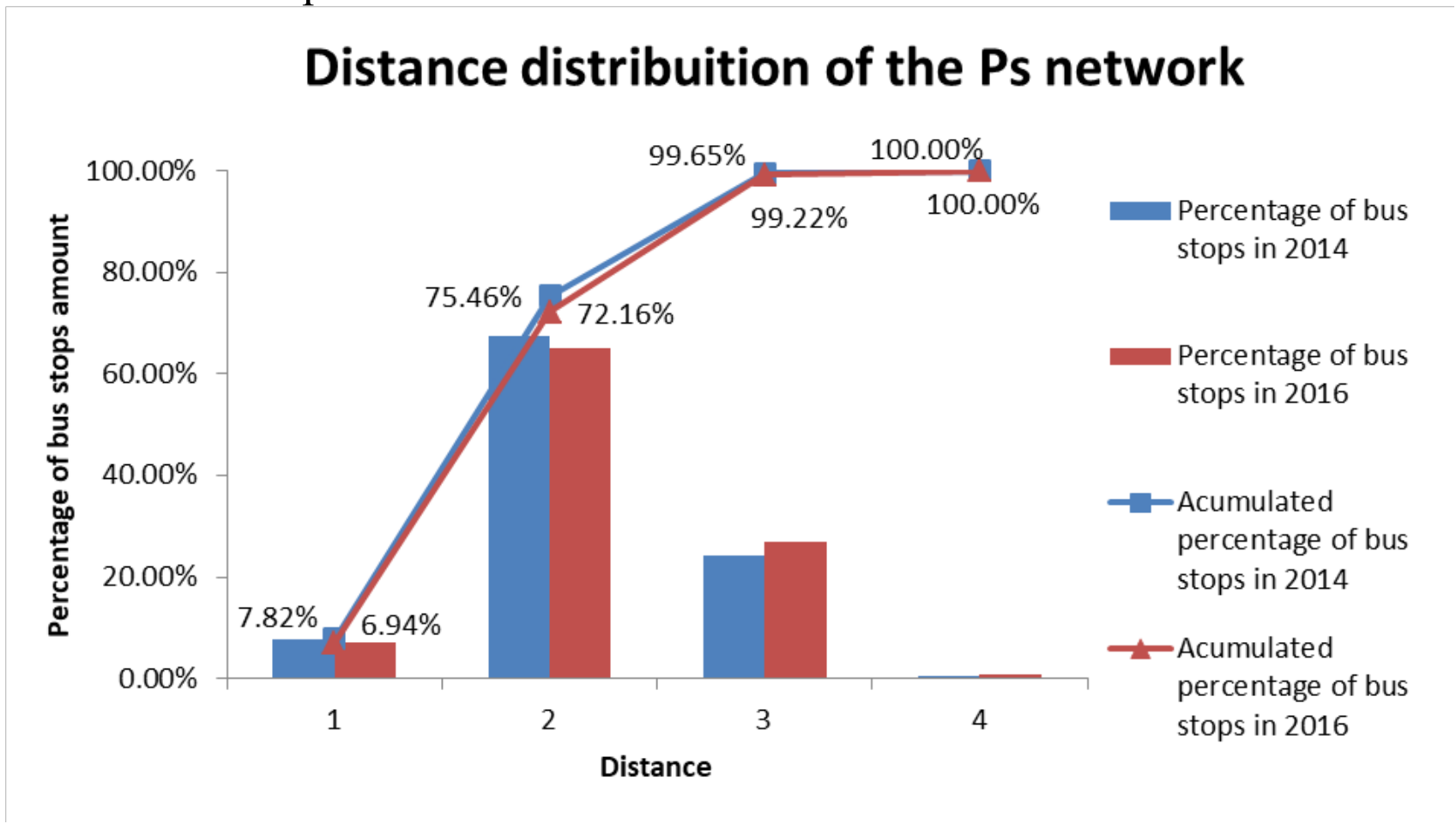

**Fig. 4.** Distance distribution of the Ps network in 2014 and 2016

The giant component in the Ps and Cs networks presents the number of nodes that are connected. In this study case, 100% of the nodes belong to the giant component. It means that all bus stops in the city are connected. Therefore, it is possible to go from any bus stop to another in the city because does not exist any line that is not connected to another at some point. This measure for the $Cs^n$ network still indicates something else. In the $Cs^{100}$ network, the giant component evinces the routes that are similar because in this kind of network, it is considered only lines that have more than 100 points in common with other lines. In 2014, the giant component had 53 routes. This means that 10.86% of the total number of lines had a similar structure. In a practical way, it would be interesting to identify these routes to study the possibility of replacing part of them by expressway corridors. In 2016, there were 25 nodes in the giant component of $Cs^{100}$ network, representing 6.63% of the lines. This indicates that there was a reduction of similar routes.

Fig. 5a shows bus stops of the lines belonging to the giant component of $Cs^{100}$ network in 2014, according to their location on the Rio de Janeiro's map. Likewise, Fig. 5b exhibits this data in 2016. Fig. 5c displays the BRT (Bus Rapid Transit) system that has been deployed in the city since 2015. In this figure, the red line is Transcarioca, which connects Barra da Tijuca neighborhood to Tom Jobim International Airport; the burgundy line is Transolímpica, which connects Barra da Tijuca and Recreio dos Bandeirantes neighborhoods to Deodoro and Magalhães Bastos; the violet line is Transoeste, which connects the West Zone to Barra da Tijuca. The yellow line is Transbrasil, which will run along Avenida Brasil and it is not finished yet.

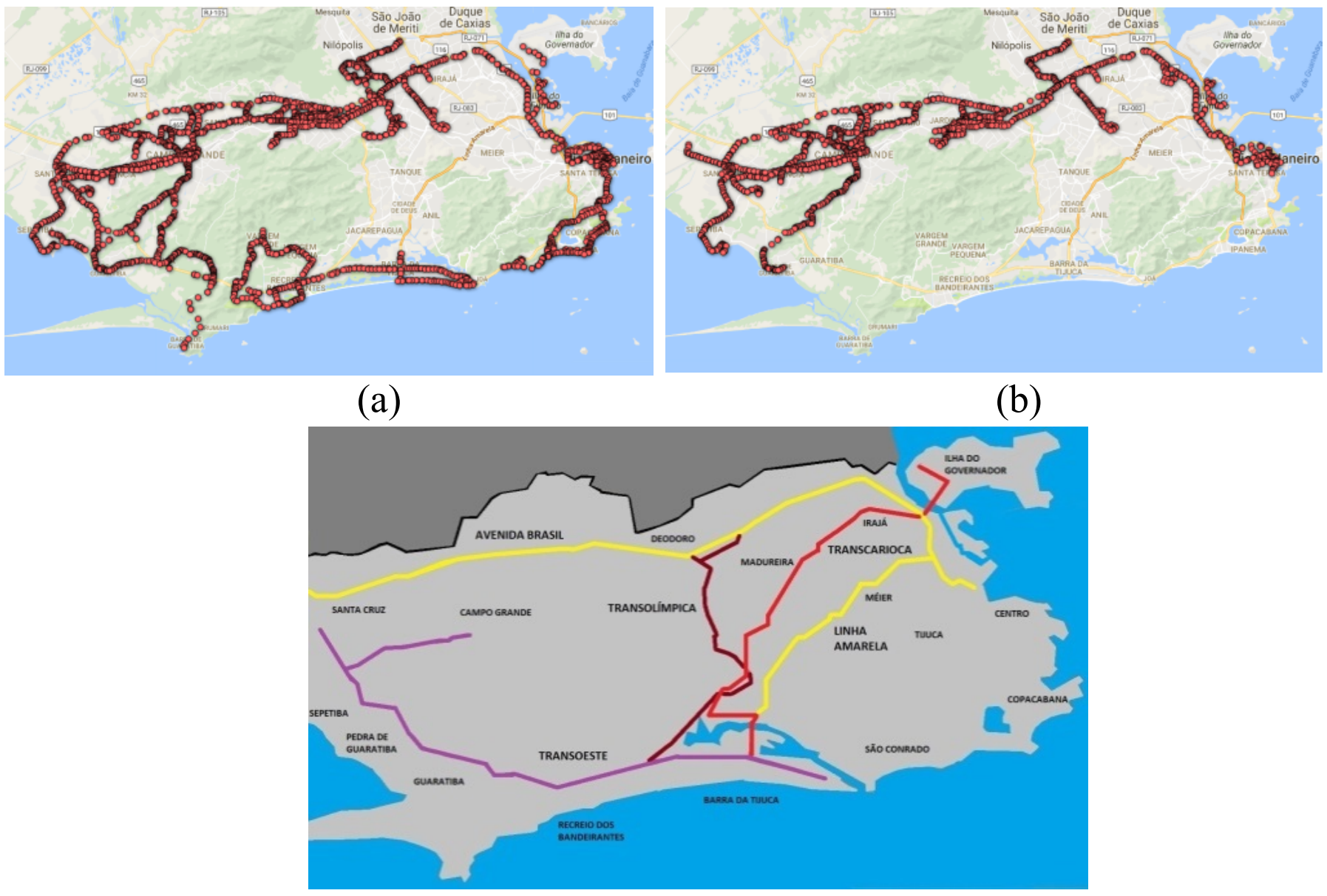

**Fig. 5.** (a) Bus stops that belong to the $Cs^{100}$ network giant component in 2014. (b) Bus stops that belong to the $Cs^{100}$ network giant component in 2016. (c) BRT system in Rio de Janeiro

When comparing Fig. 5a with Fig. 5b, there is a large reduction of the bus stops from 2014 to 2016, especially on the bottom of the map. Confronting these figures with the BRT system, it is noticed that decreased points occurred in Avenida das Américas and South Zone, where BRT's Transoeste line and BRS system were implanted, respectively. In addition, BRT Transbrasil may reduce similar lines along Avenida Brasil, since there are several routes with bus stops at this avenue.

Therefore, some expressway corridors' location created by BRT system is in the same path of some bus stops of CS100 network. Thus, it is possible to infer that the giant component propriety with many common points can be a useful methodology to verify possible expressway corridors.

As the Bs is a bipartite network, it has two average degrees. The average degree for routes partition was 110.37 in 2014 and 106.66 in 2016. This represents the average number of bus stops that a line has. The average degree for bus stops was 7.67 in 2014 and 6.04 in 2016. It shows the average amount of lines passing through a bus stop.

In the bus stops partition of Bs network, twenty bus stops with the highest degree were selected in 2014 and 2016. These points were plotted on the Rio de Janeiro's map to verify its location. The result is shown in Fig. 6. The highlight point in green represents the bus stop with highest degree. It turns out that there were some bus stops in 2014 that were not in 2016. These points were located in Terminal Alvorada and are not present in 2016 due to the implementation of BRT in this area. In addition, the map showed some points in Avenida Brasil on the Guadalupe and Caju's neighbor-hoods. However, most bus stops are located in the center of the city, at the intersection of Avenida Presidente Vargas and Avenida Francisco Bicalho. This area deserves special attention and it is presented in Fig. 7. It is observed that this area has a large number of bus stops, besides having the bus stop with the highest degree in both years.

In 2014, the twenty bus stops with highest degree had this propriety fluctuating between 49 and 78; while in 2016, ranged between 43 and 68. In 2014, the bus stop with highest degree had its value equal to 78, which means that at this single point passed 78 different lines. In 2016, this value decreased to 68. Although these values are still high, there was a degree reduction, which favors buses better traffic in this area.

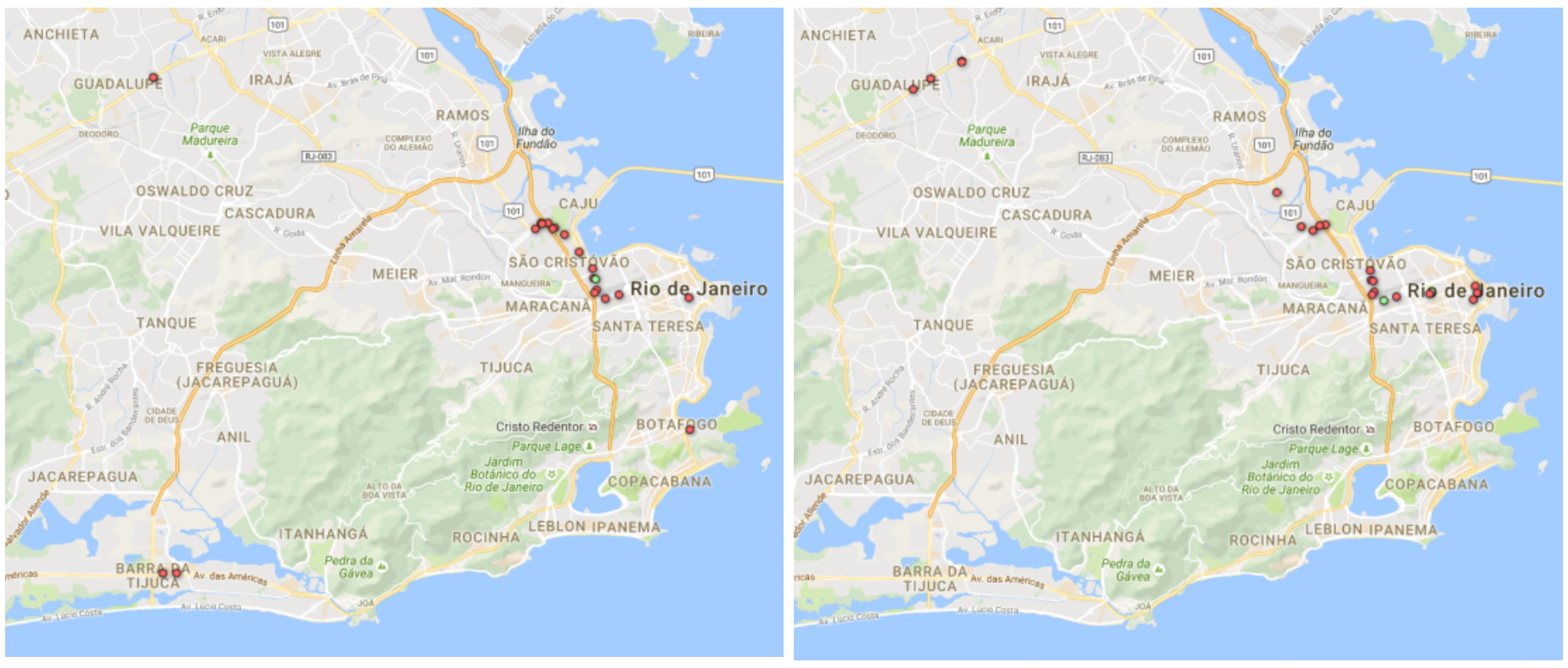

(a) (b)

**Fig. 6.** Bus stops with higher degree in *Bs* network (a) in 2014 and (b) in 2016

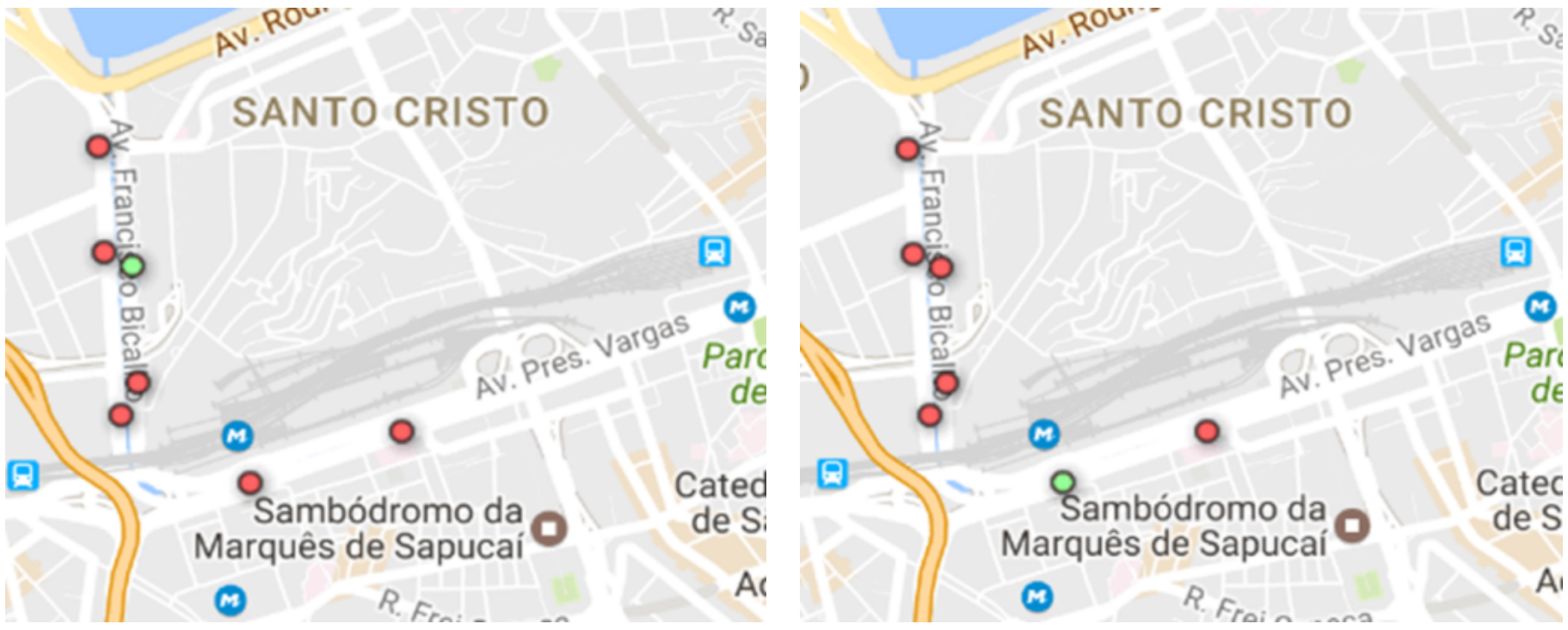

(a) (b)

**Fig. 7.** Bus stops with higher degree in *Bs* network in the downtown (a) in 2014 and (b) in 2016

Considering the line degree partition, Fig. 8 shows degree distribution, by range, of the percentage of lines in the Bs network. In 2014, the line with the highest degree stopped at 416 bus stops along its route. However, in 2016, that same line stopped at 360 points.

In general, there was a reduction of lines' degree, including the maximum degree. This decreased allows lines to be faster, since they need to stop at fewer points.

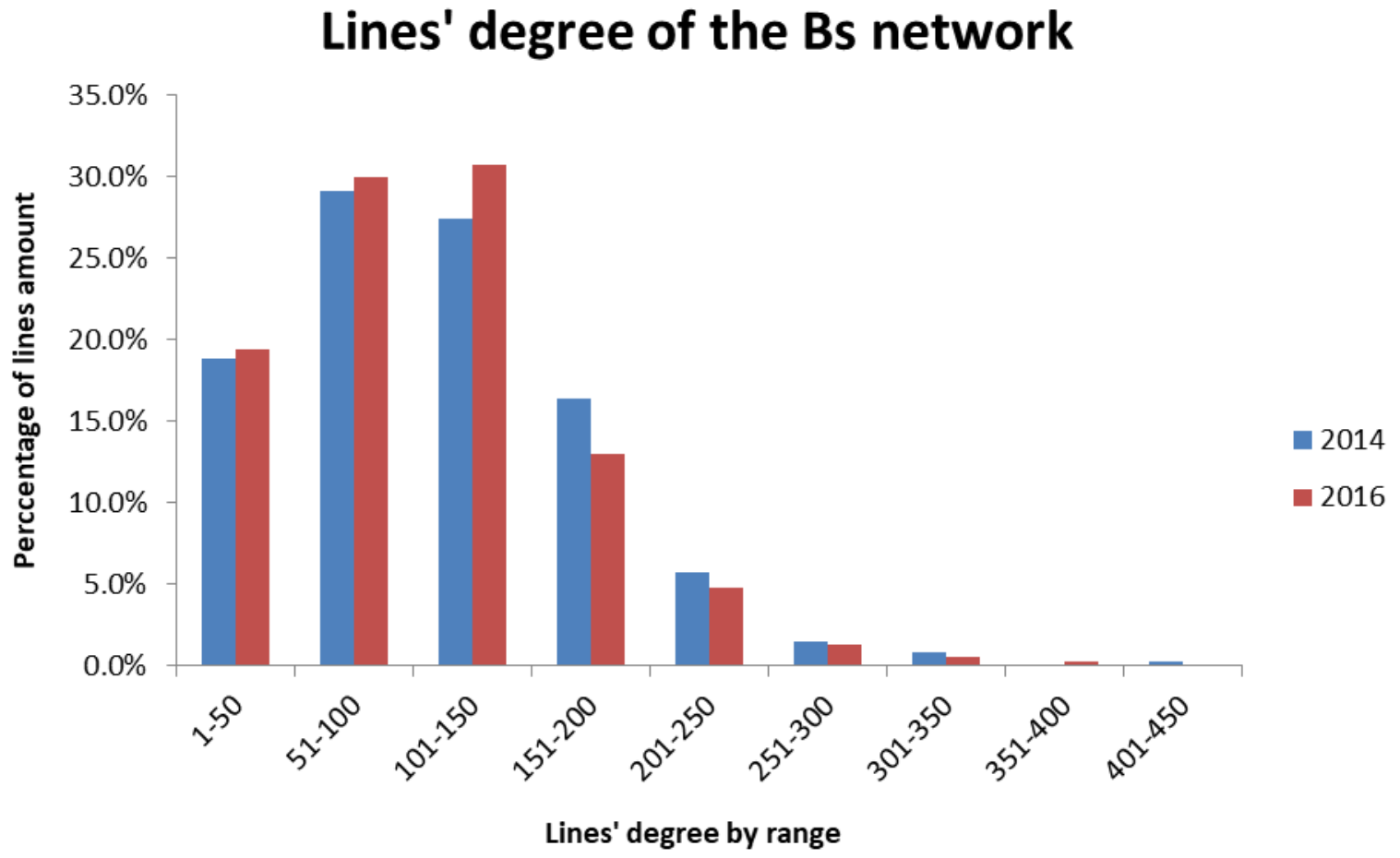

**Fig. 8.** Degree distribution of Bs network routes, by range

For the Ps network, three centrality measures were analyzed – degree, closeness and betweenness. Degree of Ps network reveals, from a certain point, how many bus stops a person can go by using only one line. In 2014, the average degree was 548, while in 2016, it was 462. This means that, given a certain bus stops, the average amount of points that a person can go decreased. The maximum level was 3239 in 2014 and 2943 in 2016. This shows that from the most central point, it is possible to go to 46% of existing bus stops; in 2016, this number fell to 44%.

Fig. 9 presents bus stops with degree higher than 2000 on the map of Rio de Janeiro in both years. The amount of bus stops decreased greatly, it was 60 in 2014 and 20 in 2016. On the other hand, location of these points remained practically the same in both years. There were a few bus stops at Avenida Brasil, by Guadalupe and Caju neighborhoods, with only seven of these points remaining on that road. Also, there was a concentration of bus stops in Cascadura, near to Madureira and Cascadura's train stations, which was reduced to only one bus stop. In addition, there was a reduction of bus stops in the city center. Besides, the same area where appears bus stops with maximum degree in Bs (Fig. 7) is also present on the map, which means that points in this area are also considered central by degree centrality in the Ps.

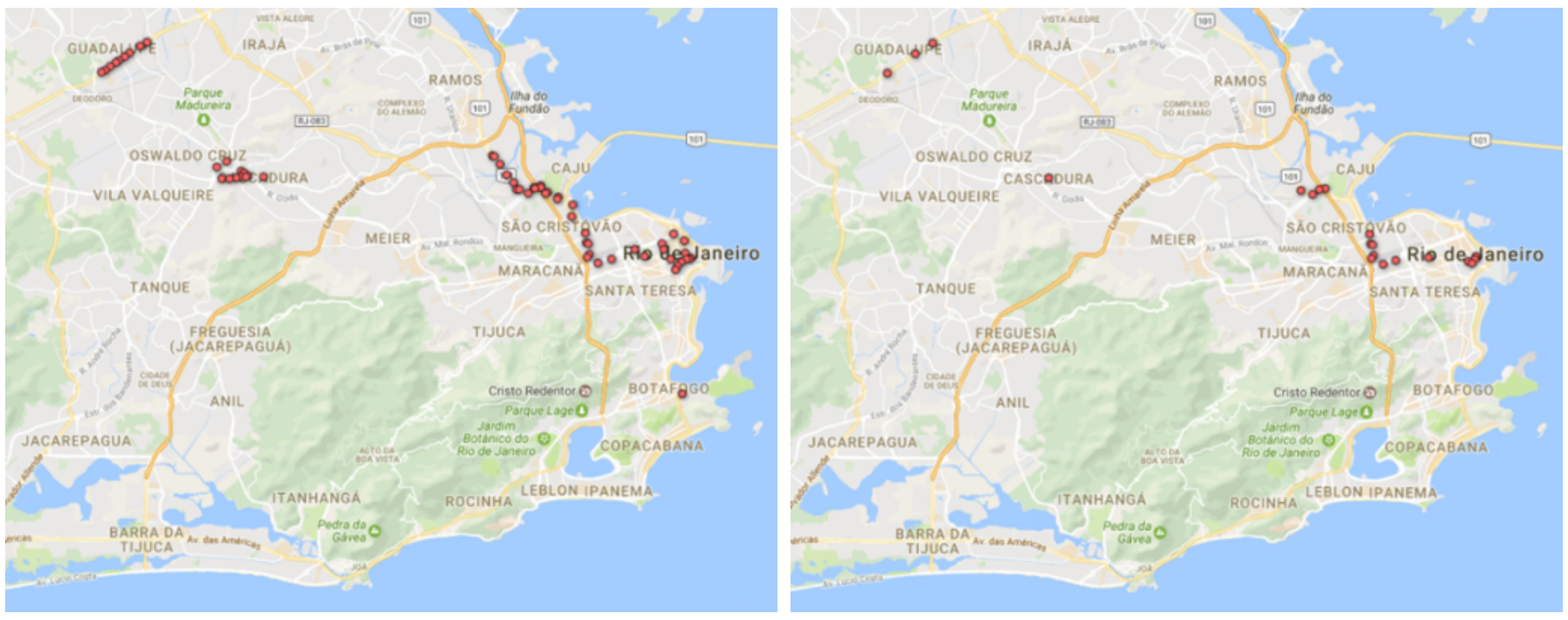

(a) (b)

**Fig. 9.** Bus stops with degree higher than 2000 in *Ps* network (a) in 2014 and (b) in 2016

Closeness centrality measure in *Ps* network indicates how close a node is to the other. If a point has high closeness value, it has smaller distance to other points. High closeness values can indicate suitable locations for possible road terminals because there may be traffic congestion at this area, since many lines pass in these points to connect them to others.

Fig. 10 exhibits bus stops with closeness higher than 0.59 on the Rio de Janeiro's map in 2014 and 2016. Bus stops amount reduced by half from 2014 to 2016. In 2014, there were 28 points and in 2016, 14 points. The bus stops location also changes a bit, becoming more grouped in 2016 than 2014. In addition, bus stops in Avenida Brasil, at Guadalupe area, which appeared in Ps and Bs' degree, did not appear in Ps' closeness propriety. Bus stops in the Cascadura area, near train station, appeared only on the map of 2014. Also, bus stops in the city center (Fig. 7) are considered central points by Ps closeness propriety in 2014 and 2016.

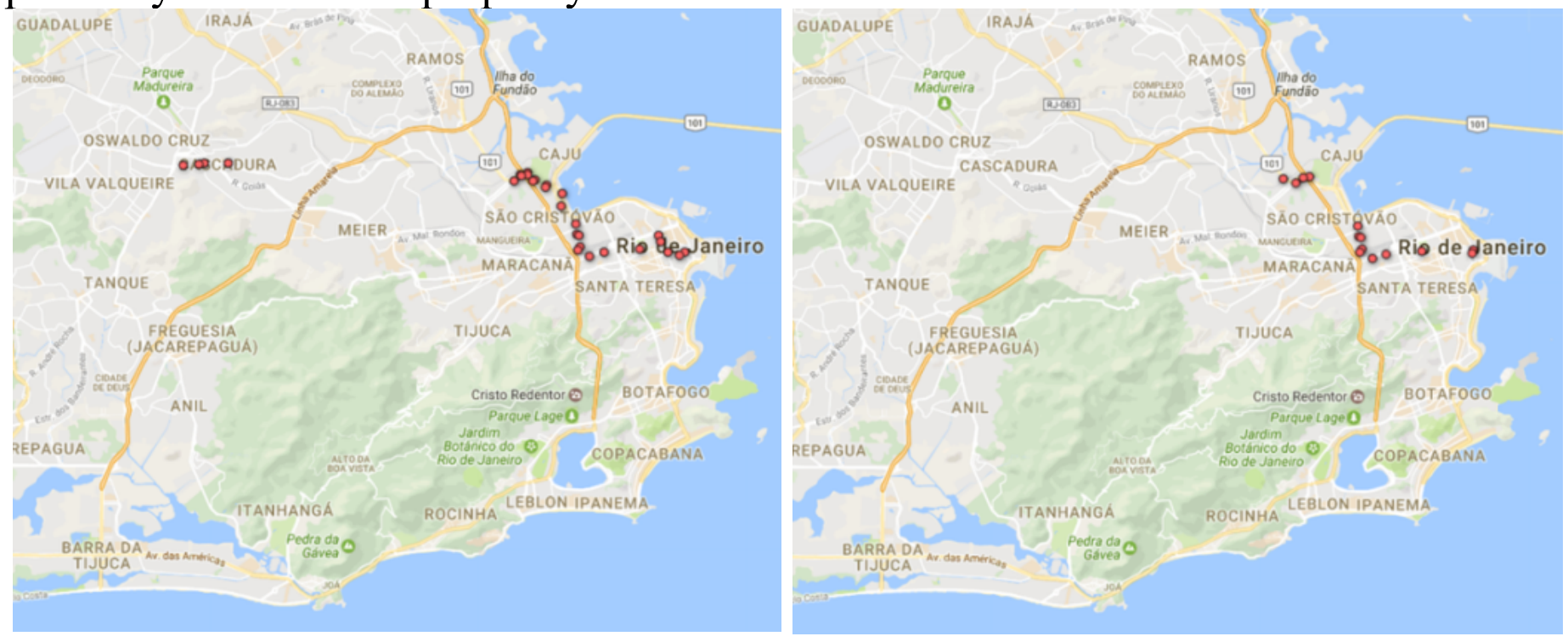

(a) (b)

**Fig. 10.** Bus stops with closeness higher than 0.59 in *Ps* network (a) in 2014 and (b) in 2016

Betweenness centrality for Ps network indicates the path amounts with shortest distances within a network that pass through a certain bus stop. Within a network, a passenger will probably pass through a bus stop that has high betweenness to go from one point to another. This happens because bus stops with high betweenness work like connectors for other points, considering the smallest possible distance. In other words, there are bus stops used for better transfer among routes.

In this work, betweenness values found were very low, close to zero. It occurs because Ps is a very dense network, with several connection points. Thus, there are many possible transfer points within this system, making betweenness values very small.

Fig. 11 shows bus stops with betweenness higher than $4.20 \times 10^{-3}$. These points amount varied little, being 21 in 2014 and 30 in 2016. This analyzed centrality propriety was the only one that increased from 2014 to 2016. It probably happens due to reduction in bus stops and routes amount – especially the lines number. So, it reduced the variety of shorter paths that connected two pairs of bus stops within the network. It made the bus stops to have a greater importance on transfer among lines.

Regarding their location, it remained practically the same in both years, with small variations in bus stops number. Some points appeared in Campo Grande, which had not been highlighted in other centrality measures. On the other hand, bus stops in Cascadura area remained as central points in this propriety. Finally, it should be noticed that bus stops in the small highlighted area in Fig. 7 also appeared in Fig. 11, which presents bus stops with betweenness above $4.20 \times 10^{-3}$ in Ps network in 2014 and 2016. Once again, it reveals the importance of these bus stops.

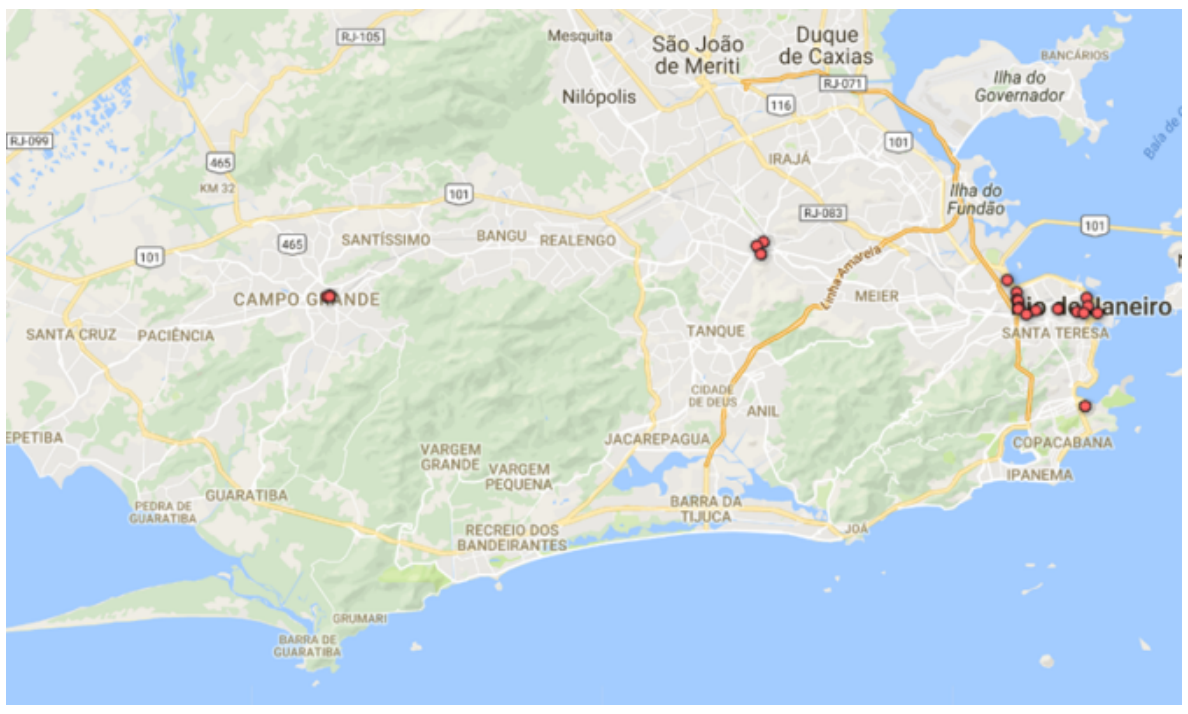

(a)

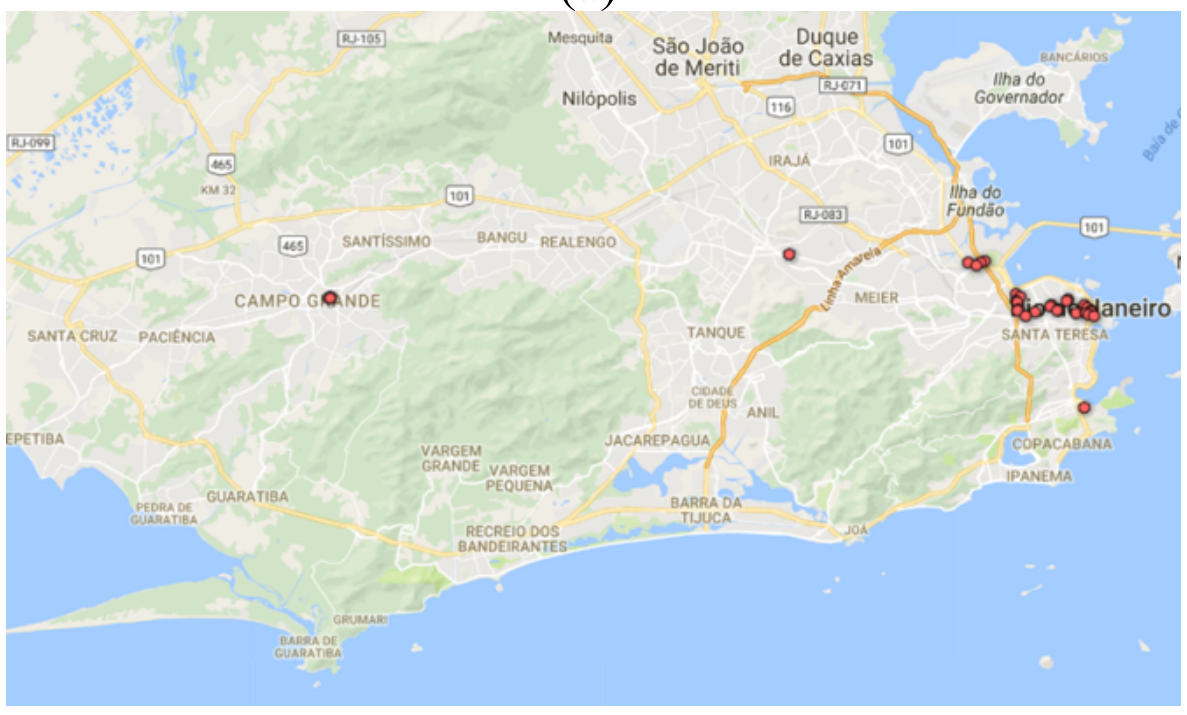

(b)

**Fig. 11.** Bus stops with betweenness higher than $4.2x10^{-3}$ in *Ps* network (a) in 2014 and (b) in 2016

For Cs network, degree centrality represents, from a certain line, the number of lines that a person can go. High degree lines can show possible expressway corridors and existence of feeder lines. Fig. 12 shows an example for better understanding of Cs network degree. Considering that buses in the figure are lines, the orange line has degree 5, the highest of the network. This line is used as transfer line among blue lines, that is, to go to one blue line to another blue line, a person will need to use the orange line. Therefore, for orange line being a much used route, this line could be replaced by an expressway corridor. It would increase its capacity and speed. Otherwise, blue routes can be considered as feeder lines of this expressway corridor because these are lines directly connected to its.

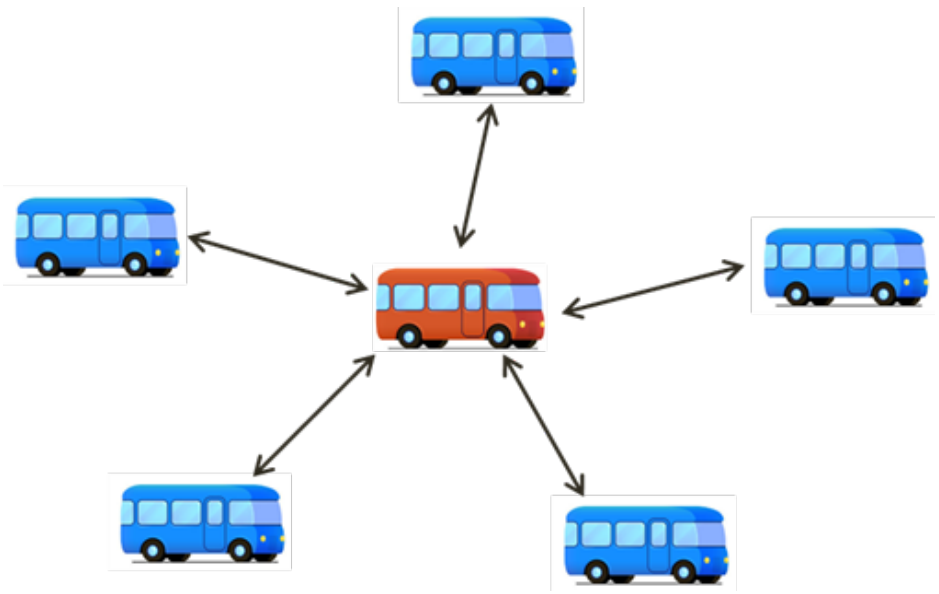

**Fig. 12.** Degree centrality explication in Cs network

In order to visualize the central lines geographical location in Cs, we selected twenty routes with highest degree and plotted bus stops belonging to these lines on the Rio de Janeiro's map. Fig. 13 and Fig. 14 show these bus stops on $Cs^{1}$, $Cs^{50}$ and $Cs^{100}$, for 2014 and 2016, respectively. As closer to red, more lines stop at this point; as closer to green, fewer lines stop there. It is noteworthy that, for a better visualization of lines intensity passing through a certain bus stop, points where only one line passed were disregarded.

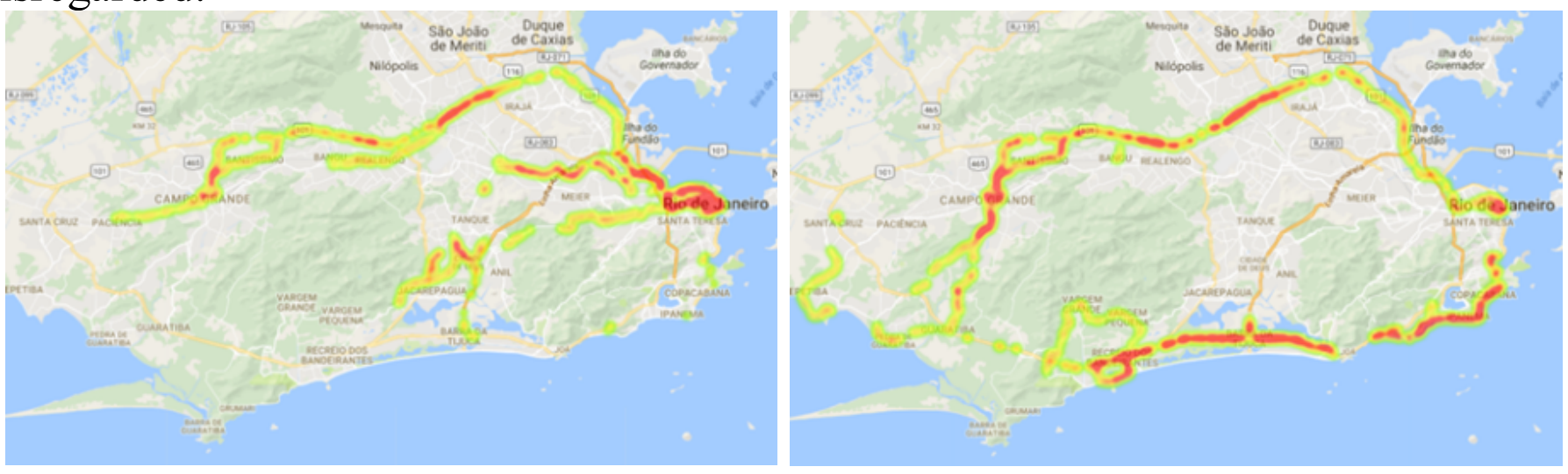

(a) (b)

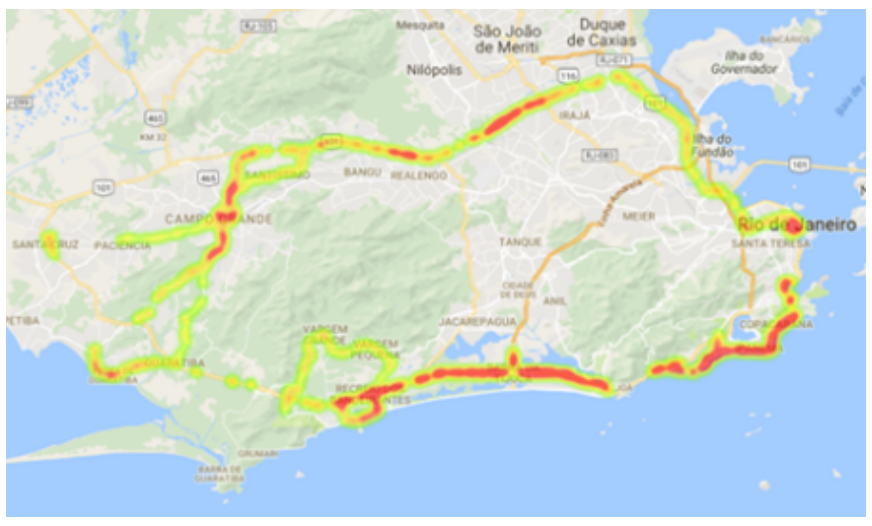

(c)

**Fig. 13.** Rio de Janeiro`s map of bus stops with higher degree in Cs network in 2014. (a) $Cs^{1}$. (b) $Cs^{50}$. (c) $Cs^{100}$.

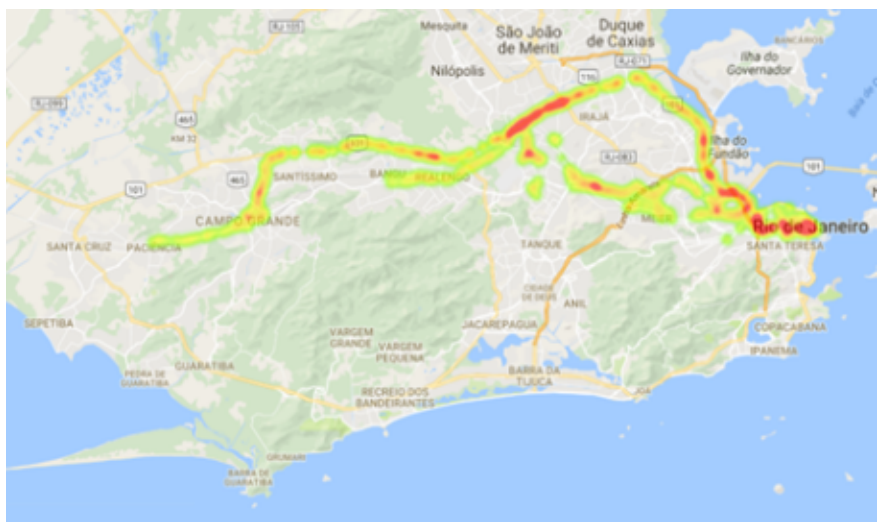
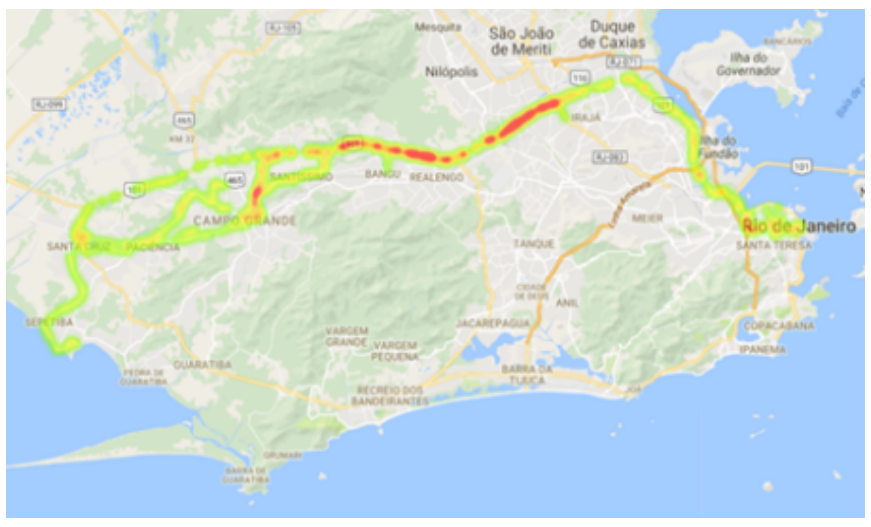

(a) (b)

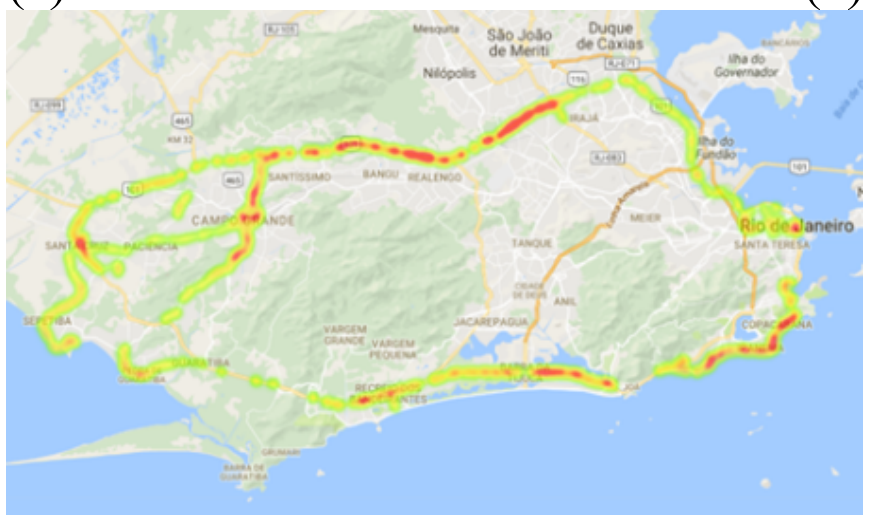

(c)

**Fig. 14.** Rio de Janeiro`s map of bus stops with higher degree in Cs network in 2016. (a) $Cs^{1}$. (b) $Cs^{50}$. (c) $Cs^{100}$.

In 2014, Cs network presents a greater concentration of bus lines passing through same points in downtown area and in a patch of Avenida Brasil. In $Cs^{50}$ and $Cs^{100}$ networks, bus stops are located especially in Avenida das Américas, in more patchs of Avenida Brasil and in South Zone's coastal area – Avenida Niemeyer, Avenida Vieira Souto and Avenida Atlântica. $Cs^{50}$ displayed more routes in red than $Cs^{100}$ network, in these mentioned areas. Also, city center still presented high degree points in its area, but in a smaller amount than $Cs^{1}$ network.

In 2016, line's degree decreased compared to 2014. The red area on the intensity map also decreased, that is, the lines amount passing through same points in 2016 was lower than in 2014. When compared to 2014, the biggest difference in Cs1, Cs50 and Cs100 networks in 2016, is the Avenida das Américas and South Zone area, which showed null or much smaller concentration of red points.

When analyzing the line with maximum degree in Cs100 network, in 2014, we verify that the line with highest degree was line 2338-CAMPO GRANDE X CASTELO (VIA ESTRADA DO MAGARCA), containing degree equal to 15. This means that this line had 100 bus stops in common with 15 different lines, evidencing the similarity of this line with the others. In 2016, the highest degree was 8, belonging to the line 358-COSMOS – CANDELÁRIA. In general, lines' degree had been reduced. This could have been because of a reduction of similar lines or a change in the bus stops that these lines stop.

Fig. 15 exhibits, in green, the bus stops of line 2338-CAMPO GRANDE X CASTELO (VIA ESTRADA DO MAGARCA) in 2014 and, in red, the bus stops of line 358-COSMOS – CANDELÁRIA in 2016. These lines pass through areas where there is a greater intensity of routes stopping at the same bus stops, such as Avenida Brasil, Avenida das Américas, South Zone and downtown city, as previously seen at the location of giant component in $Cs^{100}$ network. This is probably the reason because these lines share many bus stops with other lines.

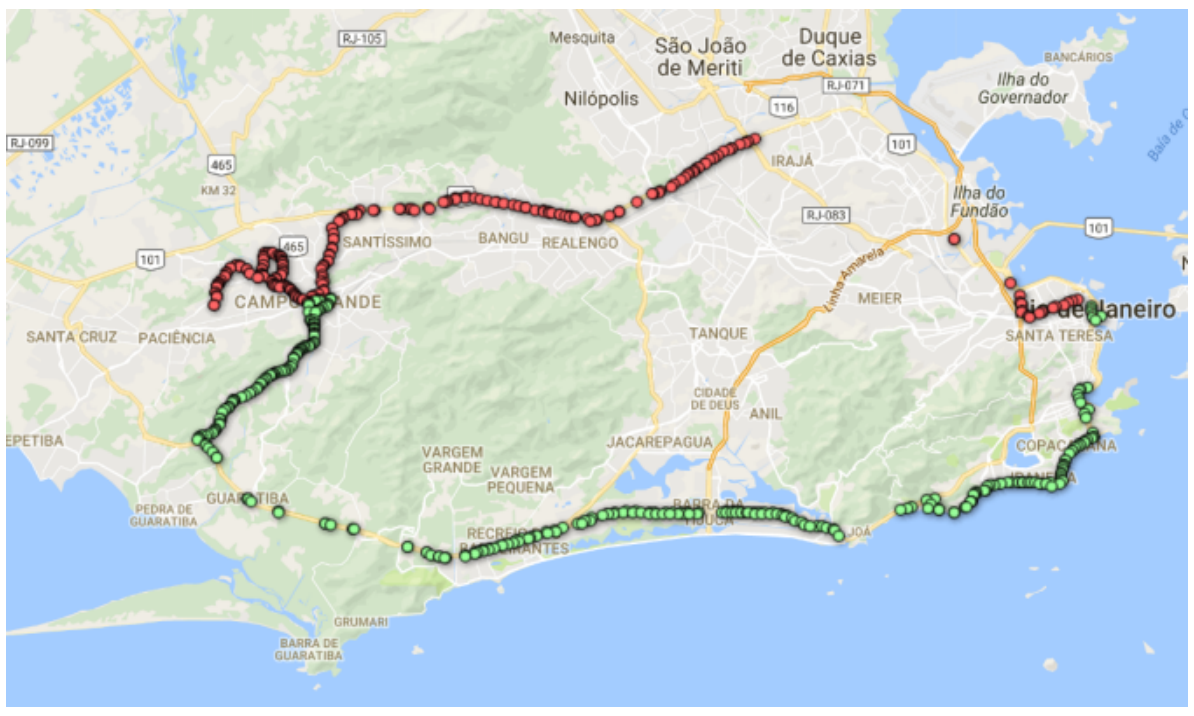

**Fig. 15.** Line 2338-CAMPO GRANDE X CASTELO (VIA ESTRADA DO MAGARCA) in 2014 (green) and line 358-COSMOS – CANDELÁRIA in 2016 (red)

Concisely, Table 1 and Table 2 present a summary of some proprieties variation for better visualization. These tables show whether propriety values have increased or decreased from 2014 to 2016. Blue down arrows means that it had reduced its value in that period. Orange up arrow means that it had increased its value from 2014 to 2016. Black dash means that value remained the same. Table 1 shows variation for Cs network, while Table 2 shows variation for Bs and Ps networks.

**Table 1.** Some proprieties variation in Cs network from 2014 to 2016

| Proprieties | $Cs^{1}$ | $Cs^{10}$ | $Cs^{20}$ | $Cs^{30}$ | $Cs^{40}$ | $Cs^{50}$ | $Cs^{60}$ | $Cs^{70}$ | $Cs^{80}$ | $Cs^{90}$ | $Cs^{100}$ |
|---|---|---|---|---|---|---|---|---|---|---|---|
| Nodes | ↓ | ↓ | ↓ | ↓ | ↓ | ↓ | ↓ | ↓ | ↓ | ↓ | ↓ |
| Average degree | ↓ | ↓ | ↓ | ↓ | ↓ | ↓ | ↓ | ↓ | ↓ | ↓ | ↓ |
| Component giant nodes | ↓ | ↓ | ↓ | ↓ | ↓ | ↓ | ↓ | ↓ | ↓ | ↓ | ↓ |
| Nodes percent- | - | ↓ | ↓ | ↓ | ↓ | ↓ | ↓ | ↓ | ↓ | ↓ | ↓ |
| age in the component giant | | | | | | | | | | | |
| Average distance | ↑ | ↑ | ↑ | ↑ | ↑ | ↑ | ↑ | ↓ | ↓ | ↑ | ↓ |
| Diameter | - | ↑ | ↑ | ↑ | ↑ | ↑ | ↑ | ↓ | ↓ | ↑ | ↓ |

**Table 2.** Some proprieties variation in Bs and Ps networks from 2014 to 2016

| Proprieties | Bs | Ps |
|---|---|---|
| Nodes | ↓ | ↓ |
| Average degree | ↓ | ↓ |
| Component giant nodes | ↓ | ↓ |
| Nodes percentage in the component giant | - | - |
| Average distance | ↑ | ↑ |
| Diameter | - | - |

## 6.1 Major results analysis

The vertices number analysis of Ps and Cs showed that the bus stops and lines' amount have shown a decrease from 2014 to 2016, resulting in a reduction of 5.19% of bus stops and 22.75% of lines. Despite this large decrease, the network topology has shown itself similar in this period, with minor modifications in the network properties. The distance property was one of those cases. There was a small increase in the distance between bus stops from 2014 to 2016, showing that even with fewer bus stops and lines, the distance was practically the same. It means that, in general, only few changes were noticed in the BTN city structure.

The three measures of centrality studied – degree, closeness and betweenness – pointed to some areas of the city of Rio de Janeiro which deserved a greater attention, since these regions hold central points by many of these measures in different networks. In other words, these are bus stops that have a high usage by people and, therefore, are considered essential points for the proper functioning of the BTN. The major area found was the city center (Fig. 7) - corner of Avenida Francisco Bicalho with Avenida Presidente Vargas – which presented several bus stops with a high centrality in all analyzed measures in the Bs and Ps networks, in both years.

Another region worth mentioning is the Cascadura and Madureira train stations, since in those areas were found some bus stops considered to be main points by measurements of degree centrality and betweenness in the Ps networks in 2014 and 2016 and also by closeness in the Ps network in 2014.

Finally, there are some bus stops in Avenida Brasil, such as Guadalupe and Caju, which stand out by measurements of degree centrality for the Bs and Ps networks in the both years.

In addition, another substantial outcome was the analysis of the Cs100 network's giant component and Cs networks' degree. It was compared to the BRT system, which was deployed in the studied period, proved to be useful properties for the identification of possible expressway corridors.

# 7 Conclusions

This work aimed to apply topological models used in Social Networks to the public transport network. It evaluated specifically the bus transport network of Rio de Janeiro city, since it had undergone a major reform of its lines and bus stops recently. For this, the topological models Bs, Ps and Cs was applied to the BTN of the city, analyzing the nodes number and the properties of giant components, distance, diameter, degree, closeness and betweenness of these networks.

Despite the large reduction in the number of lines and bus stops from 2014 to 2016, some results remained similar in both years. This was the case of the diameter in the Ps, which remained with the value of 4, showing that a maximum of 4 lines is still required to move between any two bus stops. The distance distribution of Ps also remained similar. It indicates that, on average, 2 lines continue to be needed to move between points in the city, and that 3 lines are required to access more than 99% of the bus stops.

Another important outcome was the degree. In Bs network, the degree of the line's partition showed that lines passed, on average, through 110 bus stops in 2014 and 107 in 2016. The largest line had 416 points in 2014 and 360 in 2016. On the other hand, the degree of the points' partition identified that an average of almost 8 lines passed through each bus stop in 2014. In 2016, this number was 6. The bus stop with highest degree demonstrated that 78 lines passed through this bus stop in 2014. In 2016, it was 68 lines.

A further goal of this work was the exploration of the Cs network with a certain amount of common points. To this aim, the Cs1, Cs10, Cs20, Cs30, Cs40, Cs50, Cs60, Cs70, Cs80, Cs90 and Cs100 networks were created. First, we created the $Cs^1$, the original network. Based on it, we created all its variations according to the minimum number of bus stops that they had in common. For example, the Cs100 network only had connections between the lines if they had at least 100 bus stops in common.

The giant component's analysis of Cs network allowed the identification of possible expressway corridors. We confronted the data of the expressway corridors that were created recently with the giant component of the Cs network with many common points, such as Cs100 network. When plotted on the map, it displays the path where these lines with many common bus stops pass through. Therefore, the giant component's analysis of Cs network with many bus stops in common may be a suitable method to detect possible expressway corridors.

There have been some limitations in this work that remains as proposals to future works. It is suggested that this type of analysis – with the topological models that were used – be carried out for other cities, checking for the consistency of the results found. It is also proposed to build a public transport network of Rio de Janeiro that composes the already analyzed bus network with other existing modals in the city, such as subway, train, LRT (Light Rail Vehicle), bicycles, among others. Finally, it is suggested a dynamic analysis of the Rio de Janeiro's BTN or even of another city, considering the actual displacement of the population. In other words, it is advisable to study the BTN at several schedules, weekdays and even months, to see how the network structure can fluctuate throughout the day.